\documentclass[pre,twocolumn,showpacs]{revtex4}

\usepackage{graphicx}

\begin{document}

\title{Microscale swimming: The molecular dynamics approach}

\author{D. C. Rapaport}
\email{rapaport@mail.biu.ac.il}
\affiliation{Physics Department, Bar-Ilan University, Ramat-Gan 52900, Israel}

\date{May 11, 2007}

\begin{abstract}

The self-propelled motion of microscopic bodies immersed in a fluid medium is
studied using molecular dynamics simulation. The advantage of the atomistic
approach is that the detailed level of description allows complete freedom in
specifying the swimmer design and its coupling with the surrounding fluid. A
series of two-dimensional swimming bodies employing a variety of propulsion
mechanisms -- motivated by biological and microrobotic designs -- is
investigated, including the use of moving limbs, changing body shapes and fluid
jets. The swimming efficiency and the nature of the induced, time-dependent flow
fields are found to differ widely among body designs and propulsion mechanisms.

\end{abstract}


\maketitle

\section{Introduction}

There is a growing demand -- driven by nanotechnology -- to understand and
utilize nanoscale hydrodynamical phenomena, a consequence of which is the need
to achieve a capability for detailed modeling of the underlying physical
processes occurring at submicroscopic scales. There are numerous instances where
nature itself provides examples of successful design at this level, a regime
where behavior runs contrary to the familiar macroscopic world; emulating some
of the multitude of biological architectures and mechanisms may, therefore,
provide insight into the requirements for optimized engineering design.

Microrobotics has long captured the imagination \cite{fey60}, with practical
implementations only now becoming feasible. One category of microrobot with
potential medical applications employs self-propelled swimming; however, long
before such devices were envisaged, studies of the dynamics of microscopic life
forms, and the theoretical analysis of hydrodynamics at appropriately low
Reynolds numbers, $Re \approx 10^{-5}$, made it clear that the prevailing fluid
environments were unlike those encountered in more familiar circumstances (where
$Re \gg 1$) but rather resembled slow swimming in highly viscous treacle. This
is known as Stokes (or creeping) flow, and responsibility for the
counterintuitive behavior lies in the absence of inertia from the dynamics
\cite{hap83}. Microscopic creatures swim by employing mechanisms appropriate to
these conditions, such as rotating flagella, beating cilia, body deformation,
and longitudinal or transverse surface waves \cite{pur77,chi81,vog94}. While
microscale robots may need to travel faster than their target biological
organisms in the fluid medium, their motion relative to the fluid would still be
at low $Re$.

Low-$Re$ flow is amenable to theoretical analysis because the limiting form of
the Navier--Stokes equation is linear and time independent \cite{hap83}; in this
limit, the motion of self-propelled bodies can be studied using suitable
approximations, perturbation methods and simple models, e.g.,
\cite{sha87,ehl96,sto96,bec03,avr04,naj04,avr05,fel06}. Microscopic swimmers
have also been synthesized experimentally \cite{dre05}. Swimming has been
studied numerically, based on a continuum fluid (with grid discretization) and
the immersed-boundary method for dealing with the interface between the fluid
and an elastic body \cite{cor04}, and mesoscopically described fluids utilizing,
e.g., the lattice-Boltzmann method, have also been used \cite{ear07}.

The present study of microswimmers is based on molecular dynamics (MD)
simulation \cite{rap04}, in which the fluid itself consists of discrete atoms
that interact directly with the swimmer. The MD approach is flexible in regard
to the level of structural detail that can be incorporated and, unlike other
approaches, hydrodynamic correlations emerge naturally. A selection of
self-propelled bodies that swim using a variety of mechanisms is investigated; a
particularly important aspect of the results is a comparison of the efficiency
of the alternative body designs and propulsion techniques. 

The MD approach is motivated and justified by its success with other
hydrodynamical problems, e.g., hexagonal flow patterns in Rayleigh--B\'enard
convection \cite{rap06} and toroidal rolls in Taylor--Couette flow
\cite{hir98,hir00}, the latter even in quantitative agreement with continuum
theory. Although large driving forces and gradients are required to compensate
for small size in order to exceed the instability threshold, MD demonstrates
continuum-like behavior in fluid layers a mere 30 atomic diameters thick. Thus
the scaling implicit in dynamic similarity extends down to the MD regime,
although there is an eventual lower size limit \cite{ver95}.

\section{Methodology}

The swimmer is immersed in a two-dimensional (2D) fluid (where computation
requirements are reduced substantially and flow visualization simplified
compared to 3D) confined to a rectangular container. The body and limbs of the
swimmer are constructed from soft-disk atoms. Each atom is placed at a
particular location relative to a reference coordinate frame attached to the
body; the locations are either fixed in the frame if they form the body, or
follow predetermined periodic paths if they belong to a limb or other moving
component of the propulsion mechanism. The atoms interact via a short-range,
repulsive interaction, $U(r) = 4 \epsilon [ (\sigma / r)^{12} - (\sigma / r)^6
]$, with range $r < r_c = 2^{1/6} \sigma$. The atom spacing in the swimmer is
$1.1 \sigma$, producing a rough boundary to oppose fluid slip; while a hindrance
at higher $Re$ where drag reduction is important, skin drag is necessary for
certain kinds of low-$Re$ propulsion. Swimmer mass is proportional to displaced
fluid area (excluding limbs); while motion is not totally inertia free, it
approaches this limit at low speeds. The container boundaries are made from
fixed atoms, spaced to produce impenetrable, nonslip walls (periodic boundaries
are unsuitable as they would allow bulk flow). Finally, there are the moving
atoms of the fluid; fluid properties such as viscosity are intrinsic to the
model and depend on the intermolecular forces, density, and temperature (in
contrast to a continuum representation where viscosity is normally an assigned
parameter).

Interactions among fluid atoms, as well as between fluid atoms and the swimmer
components and walls are computed at each MD timestep; the total force acting on
the swimmer (and torque, if needed) is evaluated from contributions of all fluid
atoms within range $r_c$ of any of the swimmer atoms. Standard MD techniques
\cite{rap04} are used in the force computations (for larger systems the task can
be parallelized), the equations of motion of the fluid atoms and swimmer
center-of-mass (c.m.) are integrated using a leapfrog algorithm, and a thermostat
is applied to prevent viscous heating.

\begin{figure}
\includegraphics[scale=0.95]{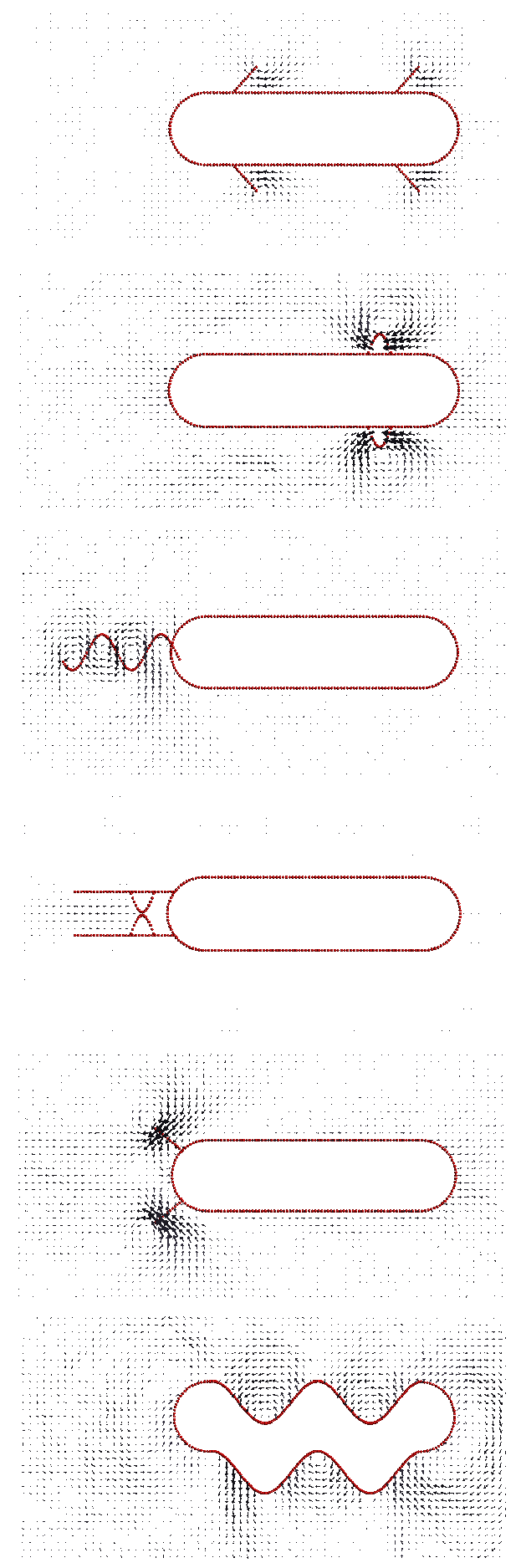}
\caption{\label{fig:f01} A selection of swimmer designs: {\em biflaps}, {\em
cigar}, {\em flagellum}, {\em jet}, {\em legs} and {\em snake}; flow fields
are shown.}
\end{figure}

A broad range of swimmer designs employing different propulsion modes are
considered; they are described below, and several are shown in
Fig.~\ref{fig:f01}. Some designs aim to emulate biological life forms manifest
at various size scales \cite{chi81,ber83,vog94}, while others implement
theoretical models (organisms that appear biologically infeasible might be
realizable artificially). Most designs are based on a cylinder -- where the term
`cylinder' refers to the 2D projection of a cigar (or spherocylinder) -- with
the same frontal area and body length to allow comparison of relative
performance. Each swimmer executes closed periodic cycles in the appropriate
internal-coordinate space; most cycles are irreversible, an essential
requirement for propulsion in the Stokes regime \cite{pur77}. Swimmer names are
indicative of shape or propulsion mechanism.

{\em Biflaps:} two pairs of lateral flaps near the front and rear rotate from
the forward direction through $90^{\circ}$ and retract. {\em Cigar:} two curved
paddles emerge laterally near the front, slide along the body to near the rear
and retract -- multiple paddles would resemble a transverse wave \cite{ehl96}.
{\em Flagellum:} a flexible rear-pointing tail along which a transverse,
two-cycle sinusoidal wave propagates -- the 2D equivalent of a rotating helix.
{\em Flaps:} a single pair of lateral flaps near the front. {\em Jet:} an open
cavity at the rear into which paddles emerge from the inner walls and slide
rearwards acting as a piston to expel the fluid, the paddles retract and there
is a recovery period for refilling the cavity -- an approximation of squid
propulsion. {\em Legs:} a pair of flaps emerges laterally at the rear, rotates
backwards through $90^{\circ}$ and retracts. {\em Slime:} a circular body that
extends a thin hollow rod (to a distance equal to the initial radius) at the end
of which a new circular body grows (at constant areal rate) while the initial
body contracts (total area is fixed), and the rod finally retracts into the new
body -- a related system is treated theoretically in \cite{avr05}.

{\em Snake:} based on a cylinder whose central axis is displaced laterally to
follow a two-cycle traveling sine wave. {\em Tail:} a rigid, rear-pointing oar
oscillates sinusoidally over a $36^{\circ}$ range; this mechanism, unlike the
others, is time-reversible, and under Stokes conditions swimming is impossible
\cite{pur77}. {\em Track:} the atoms forming the sidewalls slide along the
body, emerging from behind the front cap and disappearing into the rear, and
propulsion is due to the skin drag of the rough boundary; this is an
approximation of a membrane extruded at the front and absorbed at the rear, or
the longitudinal surface waves of cilia \cite{chi81,vog94}. {\em Trilinear:}
three linked circular bodies, joined linearly by rods whose lengths vary with
time in an irreversible cycle \cite{naj04}. {\em Tube:} hollow cylinder open to
the fluid at both ends with the same frontal area as the other cylindrical
bodies; two paddles emerge inwards, slide to the rear expelling fluid, and
retract -- a `peristaltic' mechanism.

The swimmer cylinder length is $L_b = 75$; reduced MD units are used, where the
length unit is $\sigma =$ 0.34\,nm, so that the actual length is 26\,nm.
Although this is several orders of magnitude less than even the smallest of
organisms, the MD flow studies cited earlier indicate that this should not
adversely affect the results. Other swimmer dimensions are proportional to
$L_b$; the radius is $R_b = L_b / 6$ (also the maximum radius of {\em slime}),
so the total body length is $4 L_b / 3 = 100$, flap and leg lengths are $L_b /
6$, {\em tail} oar length $L_b / 3$, {\em snake} amplitude $L_b / 10$; these and
other design elements represent reasonable (though arbitrary) choices. Careful
design is required to prevent fluid atoms becoming trapped in a corner by a
moving limb; since only the time-dependent limb position is specified, this
would cause numerical instability. Swimmer components whose size varies (such as
the growing flap) accomplish this by altering the atom overlap. The swimmer c.m.
is constrained to move in a straight line and body rotation is not allowed,
otherwise travel direction would be subject to random change (Brownian motion),
complicating the analysis.

A large container is required to help dissipate induced flows and ensure the
swimmer remains far from the walls; container size is $835 \times 338$, and for
fluid density 0.7 the total number of atoms is approximately 190\,000. Runs
begin with the swimmer at rest (except for towed bodies), positioned several
body lengths from the rear wall; the fluid atoms are assigned random velocities
with a rms value of unity, and arranged on a triangular lattice (outside the
swimmer). Run length is 610\,000 timesteps (the first 10\,000 are excluded from
analysis) of size 0.005 (MD time unit $\sqrt{m \sigma^2 / \epsilon} =$ 2.2\,ps,
where $m$ is the atom mass).

The rate of change of the internal swimmer state is governed by the drive speed
$v_d$; this corresponds to the (fixed or highest) speed of a sliding component
or the tip of a rotating limb in the swimmer reference frame. The value of $v_d$
determines the cycle period and, after allowing for interaction with the fluid,
the average swim speed and power dissipation; $v_d$ must be relatively small to
minimize density fluctuations in the (compressible) fluid and local shear rates,
but sufficiently large that runs are of reasonable duration.

Two externally towed bodies are included for comparison, moving at fixed speed
irrespective of the fluctuating fluid forces. {\em Cylinder:} the basic body
design used for most swimmers. {\em Disk:} a circular body with radius $R_b$,
but without the extra skin drag of the cylinder body.

\section{Results}

Examples of the structured, time-varying flows that develop (for $v_d = 0.6$)
appear in Fig.~\ref{fig:f01} (fluid atoms are not shown); swim direction is to
the right. The spatially coarse-grained flow is evaluated over a rectangular
grid attached to the swimmer, with velocities measured in a stationary frame;
the swim cycle is divided into 16 phase segments and each is averaged separately
over the run. Only a single segment is shown for each swimmer, but the phase
chosen is where the flow best reflects the nature of the propulsion mechanism.
Only the fluid region close to the swimmer is included to ensure adequate
resolution; furthermore, the longest arrow for each swimmer corresponds to the
highest flow speed anywhere in its cycle, and may not appear in the selected
phase.

\begin{figure}
\includegraphics[scale=0.62]{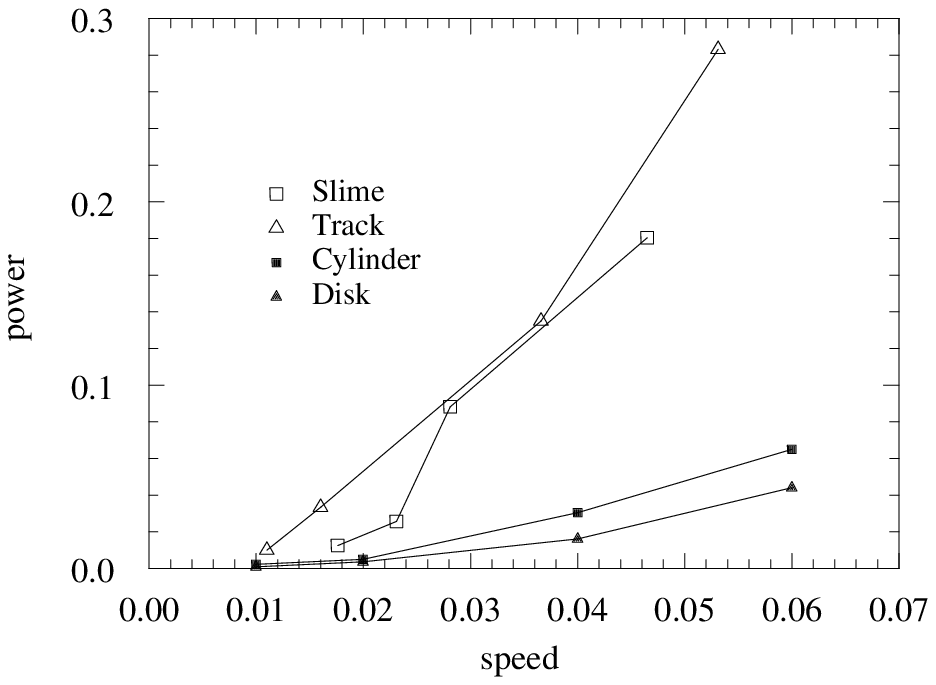}
\includegraphics[scale=0.62]{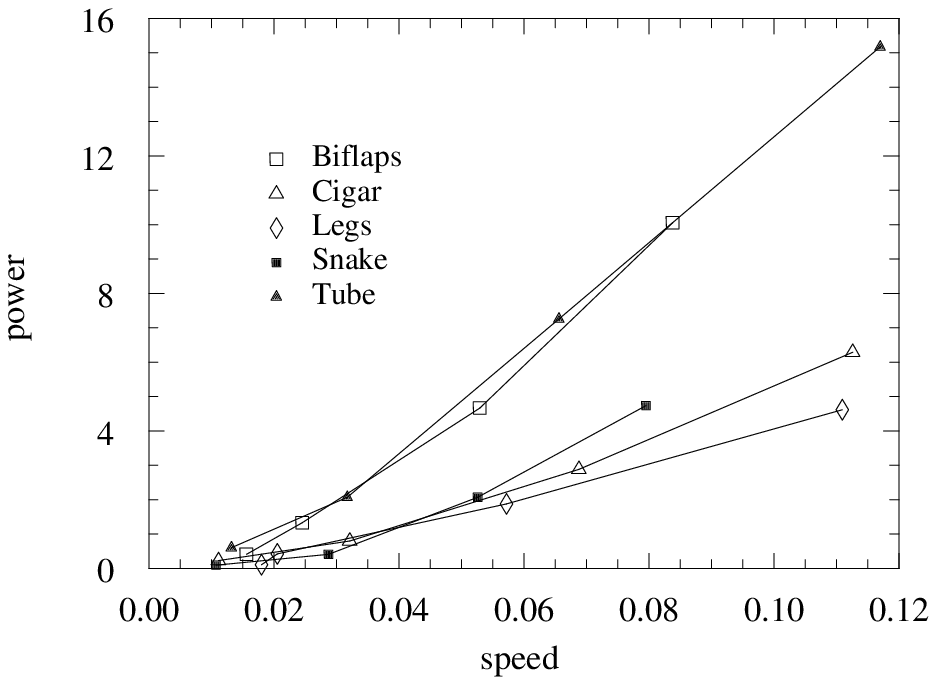}
\caption{\label{fig:f02} Cycle-averaged power as a function of swim speed for
different swimmers and drive speeds (MD units); separate graphs (with different
scales) show (a) the most efficient of the swimmers and the towed bodies, and
(b) swimmers of moderate efficiency.}
\end{figure}

Fig.~\ref{fig:f02} shows the average power $\langle P \rangle$ required to
overcome fluid resistance and maintain a mean swim speed $\langle v_s \rangle$
for several of the swimmers; the points on each performance curve are for
different $v_d$ or towing speed (in most cases 0.1, 0.2, 0.4 and 0.6, but
reduced by factors of 2, 4 and 10 for {\em slime}, {\em track} and the towed
bodies, respectively). $P$ is the total work per unit time performed by swimmer
atoms (both in the body and the propulsion mechanism) against the forces exerted
by fluid atoms within range $r_c$. $\langle v_s \rangle$ is small relative to
the fluid thermal velocity. The results are consistent with $\langle v_s
\rangle$ being proportional to $v_d$, a consequence of the linearity of the flow
problem.

Higher efficiency implies reduced power for a given swim speed; the ratio
$\langle P \rangle / \langle v_s \rangle$ provides a measure of performance --
the lower the better. Efficiency differs considerably among swimmer designs, and
for a given $v_d$, some are faster (possibly utilizing more power) than others.
Relative to the towed reference {\em cylinder} (which provides a measure of the
useful work done), {\em track} achieves a comparatively high efficiency of about
17\%, as well as an effective coupling of the body wall to the fluid since
$\langle v_s \rangle \approx v_d / 3$, while for {\em cigar} efficiency is 2\%,
and only 1\% for {\em tube}; such results lie in the expected range
\cite{sto96}. The remaining swimmers either have lower efficiency or move even
more slowly (briefly, for $v_d = 0.6$, {\em flaps}, {\em flagellum} and {\em
tail} have $\langle v_s \rangle = 0.04 \pm 0.01$ at $\langle P \rangle \approx
5$, {\em jet} has $\langle v_s \rangle = 0.04$ at $\langle P \rangle = 11$,
while {\em trilinear} is even slower and less efficient). It should be noted
that these comparisons are applicable to the specific swimmers described here;
the designs have not been optimized for maximum performance.

Establishing the relevance of the MD approach requires an estimate of $Re$. For
the midrange speed $\langle v_s \rangle = 0.04$, and a measured kinematic
viscosity $\nu = 2.1$ (from an MD analysis of 2D pipe flow -- not shown), $Re =
2 R_b \langle v_s \rangle / \nu \approx 0.5$, a value where inertial effects can
be disregarded \cite{bat67}. Since the limb speeds are higher, inertial effects
might be present at smaller length scales (explaining the motion of {\em tail}).
The towed-body data can be fit to $\langle P \rangle = C v_s^2$, the functional
form of simple Stokes drag, with $C_{disk} = 11.5$ and $C_{cylinder} = 18$ ($<
0.003$ deviation); the latter is larger due to skin friction of the elongated
body, in addition to pressure drag common to both.

\begin{figure}
\includegraphics[scale=0.62]{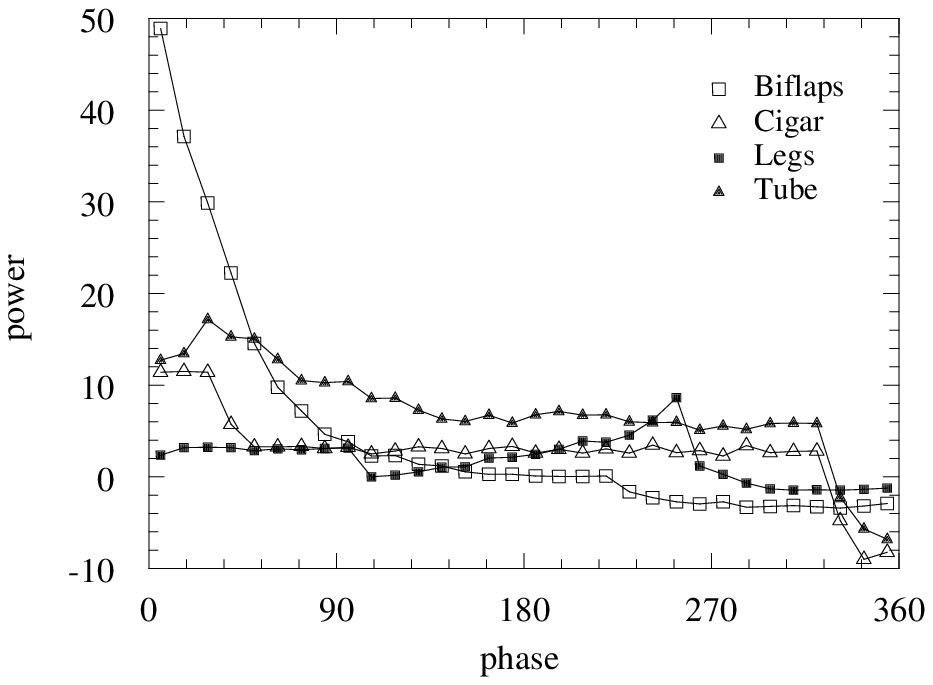}
\includegraphics[scale=0.62]{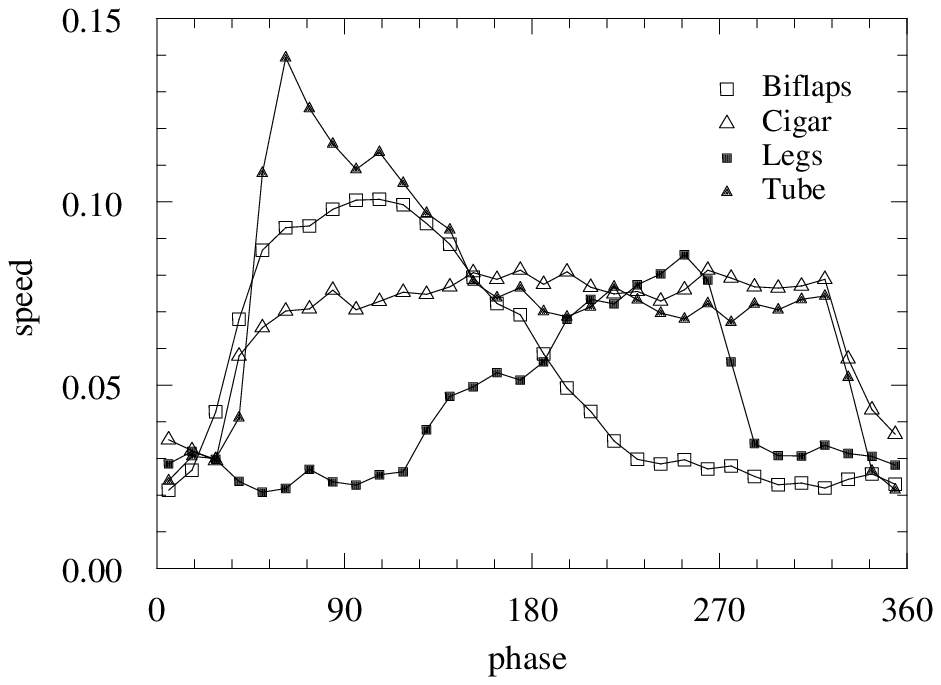}
\caption{\label{fig:f03} Dependence of (a) instantaneous power and (b) swim
speed on phase angle.}
\end{figure}

Fig.~\ref{fig:f03} shows the dependence of the instantaneous speed and power,
$v_s$ and $P$, on phase angle, for $v_d = 0.4$; the results are averaged over
the corresponding phases of all cycles in the run (15--35, depending on
swimmer). Peak values can be much larger than average, a significant design
issue, while the low $v_s$ during the phase when propulsion is absent shows that
inertial effects are small ($P < 0$ corresponds to deceleration). Each swimmer
has its own characteristic behavior reflecting the mode of propulsion. In the
examples shown, high power is needed by {\em biflaps} early in the cycle to
build up speed that then drops steadily, {\em tube} has an even more sharply
peaked early speed, for {\em legs} the increase in power and speed occurs as the
flaps approach the rear, while the values are almost constant during the sliding
portion of the {\em cigar} cycle.

In conclusion, the capabilities of a variety of swimmer designs have been
analyzed  using an MD approach built on particle-based swimmers and fluid. The
results demonstrate that MD is a viable tool for modeling swimming on the
microscopic scale and, more specifically, that swimming efficiency varies widely
among swimmer body designs and propulsion mechanisms. Since MD is limited only
by computational resources, it has the potential to play an important role in
the study of microrobotic swimmers, a field holding enormous industrial and
therapeutic promise.

\bibliography{micswim}

\end{document}